\documentstyle[12pt]{article}
\font\titlefont=cmbx10 scaled \magstep2
\begin{document}
\input{epsf}

\begin{flushright}
\vspace*{-2cm}
quant-ph/9804055 \\ 
April 22, 1998 \\
Revised Oct. 22, 1998
\vspace*{1cm}
\end{flushright}

\begin{center}
{\titlefont Casimir Force between a Dielectric Sphere and a Wall: \\
\vskip 0.2in
A Model for Amplification of Vacuum Fluctuations  }
\vskip .4in
L.H. Ford\footnote{email: ford@cosmos2.phy.tufts.edu} \\
\vskip .2in
Institute of Cosmology\\
Department of Physics and Astronomy\\
Tufts University\\
Medford, Massachusetts 02155\\
\end{center}

\vskip .3in

\begin{abstract}
The interaction between a polarizable particle and a 
reflecting wall is examined. A macroscopic approach is adopted in which the
averaged force is computed from the Maxwell stress tensor. The particular case
of a perfectly reflecting wall and a sphere with a dielectric function given by 
the Drude model is examined in detail. It is found that the force can be
expressed as the sum of a monotonically decaying function of position and of
an oscillatory piece. 
At large separations, the oscillatory piece is the dominant
contribution, and is much larger than the Casimir-Polder interaction that
arises in the limit that the sphere is a perfect conductor. It is argued that 
this enhancement of the force can be interpreted in terms of the
frequency spectrum of vacuum fluctuations. In the limit of a perfectly 
conducting sphere, there are cancellations between different parts of the 
spectrum which no longer occur as completely in the case of a sphere
with frequency dependent polarizability.  Estimates of
the magnitude of the oscillatory component of the force suggest that it may
be large enough to be observable. 
\end{abstract}
\vspace{0.6cm}
PACS categories:   12.20.Ds, 77.20.+e, 03.70.+k
\newpage

\baselineskip=14pt

\section{Introduction}
\label{sec:intro}

It was noted some time ago that if one wishes to assign a frequency spectrum
to the Casimir force between reflecting planar boundaries, the result is a
wildly oscillating function of frequency \cite{Ford88,Hacyan}. 
The integral of this function over all frequencies can only be performed 
with the aid of a suitable convergence 
factor. The net Casimir energy is much smaller than the contribution of each
individual oscillation peak. The effect of integration over all frequencies is
 almost, but not quite completely, to cancel the various frequency regions
against one another. This leads to the speculation \cite{Ford93} 
that one might be able to
upset this cancellation in some way, and thereby greatly amplify the magnitude 
of the Casimir force, and possibly change its sign. 

In the case of parallel plane boundaries, no natural way to do this has been
demonstrated. However, the Casimir-Polder interaction between a polarizable
particle and a reflecting plane offers similar possibilities. Casimir and 
Polder \cite{CP} originally derived the potential between an atom in 
its ground state
and a perfectly reflecting wall. In the large distance limit, their result
takes the particularly simple form\footnote{Gaussian units with $c=\hbar=1$
will be used in this paper.}
\begin{equation}
V_{CP} \sim - \frac{3\, \alpha_0}{8 \pi\, z^4}\,,   \label{eq:CP}
\end{equation}
where $z$ is the distance to the wall, and $\alpha_0$ is the static 
polarizability of the atom. This asymptotic potential may be derived from
the interaction Hamiltonian
\begin{equation}
H_{int} = - \frac{1}{2}\, \alpha_0\, {\bf E}^2 \,, \label{eq:H_int}
\end{equation}
where ${\bf E}$ is the quantized electromagnetic field operator. If one expands
this operator in terms of a complete set of the Maxwell equations in the presence
of the boundary, the asymptotic Casimir-Polder may be written as
\begin{equation}
\langle H_{int} \rangle = \frac{\alpha_0}{4 \pi \, z^3}\, 
\int_0^\infty d \omega\, \sigma(\omega) \,,
\end{equation}
where 
\begin{equation}
\sigma(\omega) =  \Bigl[(2\,\omega^2\,z^2 -1) \sin 2\omega z
+ 2\,\omega\,z \cos 2\omega z \Bigr] \,.
\end{equation}
The integrand, $\sigma(\omega)$,
 is an oscillatory function whose amplitude {\it increases}
with increasing frequency. Nonetheless,
the integral can be performed using a convergence factor ({\it e.g.}, insert a
factor of $e^{-\beta\, \omega}$ and then let $\beta \rightarrow 0$ after
integration). The result is the right hand side of Eq.~(\ref{eq:CP}).
It is clear that massive cancellations have occurred (see Fig. 1), and that 
the area under an oscillation peak can be much greater in magnitude than
the final result. This again raises the possibility of tampering with this
delicate cancellation, and dramatically altering the magnitude and sign of the
force.

\begin{figure}
\begin{center}
\leavevmode\epsfysize=6cm\epsffile{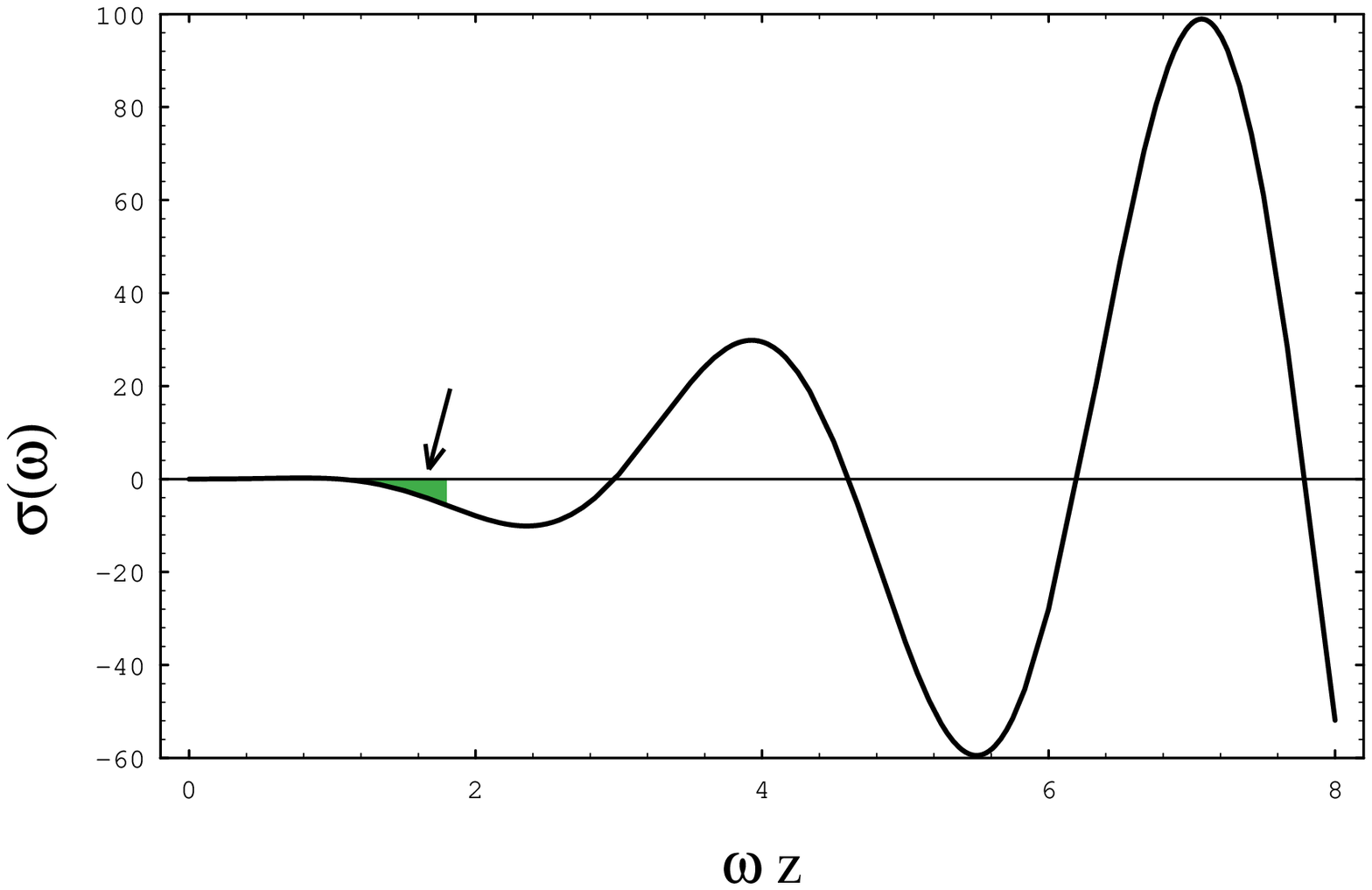}
\label{Figure 1}
\end{center}
\begin{caption}[]

 The frequency spectrum, $\sigma(\omega)$, for the Casimir-Polder
potential.  The oscillations almost exactly cancel, leaving
a net area under the curve equal to that of the shaded region indicated
by the arrow. 
\end{caption}
\end{figure}

The purpose of this paper is to explore this question in the context of a
specific model. The force between a dielectric sphere and a perfectly
conducting plane will be examined. The polarizability of the sphere will
be taken to be a function of frequency, thereby introducing the possibility
of modifying the contributions of different parts of the spectrum. This or
similar problems have been discussed before by several authors. However, it 
will here be examined from a different viewpoint. 
The force may be calculated from the Maxwell stress tensor.
In Section ~\ref{sec:dipole}, a formula for the force on a small sphere 
in an arbitrary applied electromagnetic field will be derived in an electric 
dipole approximation and discussed. In Section ~\ref{sec:interface}, 
this formula will be applied to
the calculation of the force on a dielectric particle near an interface
in terms of the Fresnel coefficients of the interface. This result will be 
applied to the case of a dielectric sphere and a perfectly reflecting boundary 
in Section ~\ref{sec:perfect}. It will be shown that the force has a component 
which is an oscillatory function of position, and that it is possible for the 
sphere is be in stable equilibrium at a finite distance from the boundary.
The results are summarized and discussed in Section~\ref{sec:final}.

\section{Force on a Small Particle}

\subsection{Electric Dipole Approximation}
\label{sec:dipole}

In this section, we will discuss the force which an applied electromagnetic 
field exerts on a small dielectric sphere. The applied electric field will be 
taken  to be ${\bf E}_a({\bf x},t)$, and the corresponding magnetic field 
to be ${\bf B}_a({\bf x},t)$.  We assume that the induced (scattered)
field is that of electric dipole radiation from a time-varying dipole moment 
${\bf p}$. Later, ${\bf p}$ will be taken to be linearly related to ${\bf E}_a$,
but for now it is unspecified. We further assume that the particle is small 
compared to the characteristic spatial scale over which  ${\bf E}_a({\bf x},t)$ 
and  ${\bf B}_a({\bf x},t)$ vary. The latter assumption is not really 
independent of the electric dipole approximation: If the size of the sphere
is not small then one would in general have to include the contributions of
higher multipoles. Just outside  the particle, the
electric and magnetic fields due to the dipole take the near-zone forms:
\begin{eqnarray}
{\bf E}_d &\approx& \frac{3 {\bf \hat n} ({\bf \hat n}\cdot {\bf p} - {\bf p}) } 
                       {r^3} 
                                              \nonumber \\
{\bf B}_d &\approx& - \frac{{\bf \hat n} \times {\bf \dot p} }{r^2} \,.
                                                      \label{eq:dipole}
\end{eqnarray}
Here $r$ is the radial distance from the dipole, and ${\bf \hat n}$ is the
outward directed unit normal vector.
  
The net force acting upon the particle can be calculated by integrating
the Maxwell stress tensor over a spherical surface just outside the particle,
\begin{equation}
F^i =  \oint d a_j \, T^{ij} \, ,
\end{equation}
where
\begin{equation}
T^{ij} = \frac{1}{4 \pi}\, \Bigl[ E^i\, E^j + B^i\, B^j - 
         \frac{1}{2} \delta^{ij}\,({\bf E}^2 + {\bf B}^2) \Bigr] \,.
\end{equation}
If we insert the net fields, ${\bf E}_a + {\bf E}_d$ and ${\bf B}_a + {\bf B}_d$,
into this expression, there will be three types of terms: those involving
only the applied fields, those involving only the dipole fields, and the
cross terms. However, the pure dipole terms yield no net contribution.
Furthermore, any force due to the pure applied field terms is independent
of the polarizability, and hence not of interest. Thus we consider only the
 cross terms in $T^{ij}$ between the applied and dipole fields:
\begin{equation}
{\bf F} = \frac{1}{4 \pi}\, \oint d a \,\Bigl[ 
({\bf \hat n}\cdot{\bf E}_a){\bf E}_d
+ ({\bf \hat n}\cdot{\bf E}_d){\bf E}_a +
({\bf \hat n}\cdot{\bf B}_a){\bf B}_d - 
{\bf \hat n}\bigl({\bf E}_a\cdot{\bf E}_d + 
                             {\bf B}_a\cdot{\bf B}_d\bigr) \Bigr] \,.
                            \label{eq:stress}                   
\end{equation}
Note that ${\bf \hat n}\cdot{\bf B}_d = 0$. 

Because the particle is assumed to be small, we may expand ${\bf E}_a$
and ${\bf B}_a$ in a Taylor series around ${\bf x} = {\bf x}_0$, 
the location of the particle.
The leading nonzero contributions to the force come from the zeroth order term
in ${\bf B}_a$ and the first order term in ${\bf E}_a$:
\begin{eqnarray}
{\bf B}_a({\bf x},t) &\approx& {\bf B}_0 \nonumber \\
 E_a^i({\bf x},t) &\approx&  
   E^i_0 + r \,{\bf \hat n}\cdot{\bf \nabla}E^i_0 + \cdots  \,,
\end{eqnarray}
We now insert these expansions and Eq.~(\ref{eq:dipole}) into 
Eq.~(\ref{eq:stress}) and perform the angular integration, using the 
relation
\begin{equation}
\oint d a \, n^i \, n^j = \frac{4\pi r^2}{3} \delta^{ij} \,,
\end{equation}
 to find
\begin{equation}
F^i = \frac{2 }{3}\, p^j \,\partial_j E^i_0 + 
\frac{1}{3}\, p_j \,\partial^i E^j_0 
     + \frac{2 }{3}\, ({\bf \dot p} \times {\bf B}_0)^i \,. \label{eq:force0}
\end{equation}

It is of interest to check the static limit of this expression. In this limit,
$\dot{\bf p} = 0$ and ${\bf \nabla}\times{\bf E}_0 = 0$. If we use these
relations, and set ${\bf p} = \alpha_0 {\bf E}_0$, where $\alpha_0$ is the
static polarizability of the particle, the result is
\begin{equation}
F^i = \alpha_0 \,p_j \, \partial^i E^j_0 = 
\frac{1 }{2}\,\alpha_0 \,\partial^i {\bf E}_0^2 \,.
\end{equation}
This is equivalent to the familiar result that the interaction energy of 
an induced dipole with a static electric field is
\begin{equation}
V = -\frac{1 }{2}\,\alpha_0 \,{\bf E}_0^2 \,.
\end{equation}

\subsection{Interaction with a Single Plane Wave}
\label{sec:plane}

Here we apply the result, Eq.~(\ref{eq:force0}), to compute the force which a 
single, linearly polarized plane wave exerts on the particle. The electric
and magnetic fields of this wave are given by
\begin{eqnarray}
{\bf E}_0 &=& Re 
\Bigl(\hat{\bf \epsilon}\, A e^{i({\bf k}\cdot{\bf x}_0 -\omega t)}\Bigr)
= \hat{\bf \epsilon} A \cos({\bf k}\cdot{\bf x}_0 -\omega t) \nonumber \\
{\bf B}_0 &=& \hat{\bf k} \times \hat{\bf \epsilon}\, A
                             \cos({\bf k}\cdot{\bf x}_0 -\omega t)\,,
\end{eqnarray}
where $A$ is the amplitude and $\hat{\bf \epsilon}$ the polarization vector.
The dipole moment is given by 
\begin{equation}
{\bf p} = Re \bigl(\alpha {\bf E}_0) =
\hat{\bf \epsilon} A |\alpha| \cos({\bf k}\cdot{\bf x}_0 -\omega t +\gamma)\,,
\end{equation}
where
\begin{equation}
\alpha = |\alpha| \, e^{i\gamma} = \alpha_1 + i \alpha_2 \,.
\end{equation}
We are interested in the time-averaged force, measured over time scales long
compared to $1/\omega$; so we henceforth understand $F^i$ to be the time average
of Eq.~(\ref{eq:force0}). In the present case, this yields
\begin{equation}
{\bf F} = \frac{1}{2}\,{\bf k}\, A^2 \,|\alpha| \sin\gamma 
= \frac{1}{2}\,{\bf k}\, A^2 \,\alpha_2 \,,   \label{eq:single_plane}
\end{equation}
a force proportional to the imaginary part of the polarizability, $\alpha_2$. 
This result
may be given a simple physical interpretation. The rate at which electromagnetic
energy is dissipated is given by the usual Joule heating term
\begin{equation}
\dot W = \int {\bf J}\cdot {\bf E}\; d^3 x \,,
 \end{equation}
where ${\bf J}$ is the current density, and the integration is taken over
the volume of the particle. Because the electric field is approximately constant
over this volume, and because one may show \cite{Jackson} from the continuity
equation that
\begin{equation}
\int {\bf J}\, d^3 x = \dot{\bf p} \,,
\end{equation}
we have that the time-averaged power absorbed by the particle is
\begin{equation}
\dot W = \frac{1}{2} \omega A^2 |\alpha| \sin\gamma \,.
\end{equation}
However, each photon carries energy $\omega$ and momentum ${\bf k}$, so the
right hand side of Eq.~(\ref{eq:single_plane}) is just the rate at which
momentum is being absorbed by the particle due to the absorption of photons.
There is of course also some momentum being transferred as a result of photon
scattering. However, that effect is proportional to $\alpha^2$, and is being
neglected here.

\section{Force on a Particle near an Interface}
\label{sec:interface}

In this section, we will derive a formula for the Casimir force on a
polarizable particle in the presence of a single plane interface. The
interface will be assumed to have arbitrary reflectivity. We will, however,
work in an approximation in which evanescent modes are neglected.
The quantized electromagnetic field is to be expanded in a complete set
of normalized solutions of the Maxwell equations. These solutions fall into 
three classes:
\begin{enumerate}

\item Modes which are in the region above the interface, and which consist of an
incident and a reflected part, as illustrated in Fig. 2.

\item Modes which originate on the far side of the interface, and which are
outwardly propagating transmitted waves in the region above the interface.

\item Evanescent modes which are propagating inside of the material
comprising the interface, but which are exponentially decaying in the 
region above it. These modes will be left out of the present discussion.
\end{enumerate}

\begin{figure}
\begin{center}
\leavevmode\epsfysize=5cm\epsffile{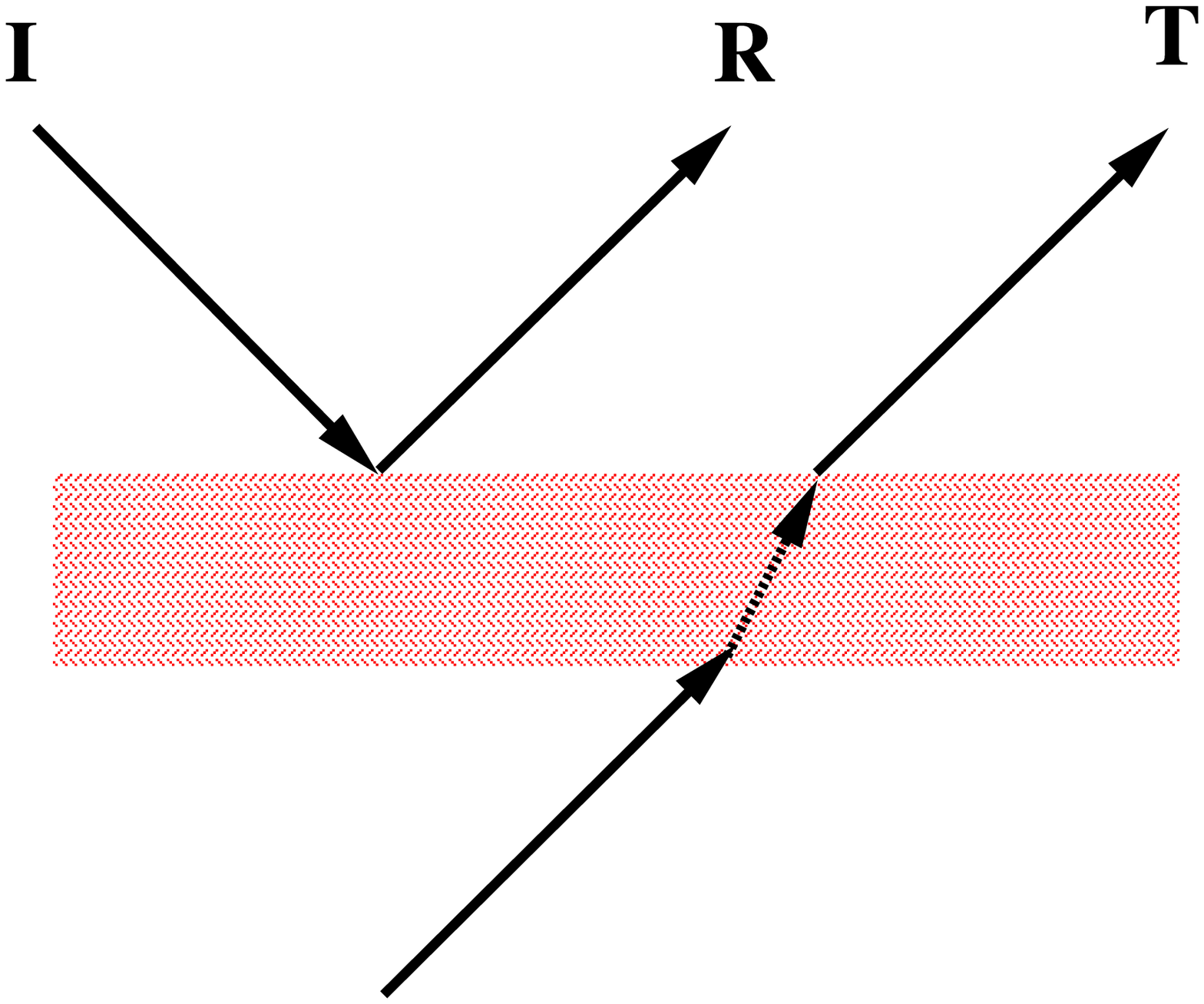}
\label{Figure 2}
\end{center}
\begin{caption}[]

The propagating modes above an interface consist of (1) incident, 
$I$,
and reflected, $R$, waves, or (2) transmitted, $T$, waves.
\end{caption}
\end{figure}

Let us focus first on the reflected modes in Class 1. The net electric field
is 
\begin{equation}
{\bf E} = {\bf E}_I + {\bf E}_R \,,
\end{equation}
where the incident wave is
\begin{equation}
{\bf E}_I = \hat{\bf \epsilon}\, A\, \cos({\bf k}\cdot{\bf x}_0 -\omega t)\,,
                                                  \label{eq:incident}
\end{equation}
and the reflected wave is 
\begin{equation}
{\bf E}_R = \hat{\bf \epsilon'}\, A\, R\,  
                    \cos({\bf k'}\cdot{\bf x}_0 -\omega t +\delta)\,.
\end{equation}
The associated magnetic fields are ${\bf B}_I = {\bf \hat k}\times{\bf E}_I$
and  ${\bf B}_R = {\bf \hat k'}\times{\bf E}_R$, respectively.
Here the complex reflection (Fresnel) coefficient is 
\begin{equation}
{\cal R} = R\, e^{i \delta} \,,
\end{equation}
where $R$ is the magnitude of the reflection coefficient, and $\delta$ is the 
phase shift. This mode induces a dipole moment ${\bf p} = Re(\alpha {\bf E})$,
where $\alpha = |\alpha|\, e^{i \gamma}$ is again the complex polarizability.
The portions of ${\bf p}$ arising from the incident and reflected waves are,
respectively,
\begin{equation}
{\bf p}_I = \hat{\bf \epsilon}\, A\,
          \cos({\bf k'}\cdot{\bf x}_0 -\omega t  +\gamma)\,.
\end{equation}
and
\begin{equation}
{\bf p}_R = \hat{\bf \epsilon}\, A\, R\, |\alpha|\, 
          \cos({\bf k}\cdot{\bf x}_0 -\omega t +\delta +\gamma)\,.
\end{equation}

The force which a particular mode exerts on the polarizable particle is 
obtained by inserting the above expressions for the fields and dipole moment
into Eq.~(\ref{eq:force0}). The resulting expression should then be summed 
over all modes. However, it is simpler first to combine it with the 
corresponding expression arising from the transmitted waves of Class 2.
In the region above the interface, the electric field of these modes is of
the form
\begin{equation}
{\bf E}_T = \hat{\bf \epsilon}\, A\, T\, \cos({\bf k}\cdot{\bf x}_0 -\omega t)\,,
\end{equation}
where $T$ is a transmission coefficient. Here we may think of the interface as
being a slab of finite thickness. Below the slab, these modes have the same form
as the incident waves above the slab,  Eq.~(\ref{eq:incident}). If the material 
in the slab is non-absorptive, then the transmission and reflection coefficients
satisfy
\begin{equation}
T^2 + R^2 = 1 \,.  \label{eq:TR}
\end{equation}
The force due to the modes of Class 1 can be expressed as a sum of three types
of terms, those involving only the incident wave, those involving only the
reflected wave, and cross terms between the two. (See Fig. 3.)
 The first two types of
contributions are of the form discussed in the previous section for a single
plane wave, as are the contributions due to the Class 2 transmitted waves.
As a consequence of the relation Eq.~(\ref{eq:TR}), these three sets of 
contributions cancel one another, leaving only the incident-reflected-wave
cross terms. The resulting force, for a single mode, is
\begin{equation}
F^i = \frac{2 }{3}\, (p_I^j\, \partial_j E_R^i + p_R^j\, \partial_j E_I^i)  
+ \frac{1}{3}\, (p_{Ij}\, \partial^i E_R^j + p_{Rj}\, \partial^i E_I^j)
 + \frac{2 }{3}\,[ (\dot{\bf p}_I \times {\bf B}_R)^i + 
(\dot{\bf p}_R \times {\bf B}_I)^i ]\,. \label{eq:force1}
\end{equation}

\begin{figure}
\begin{center}
\leavevmode\epsfysize=5cm\epsffile{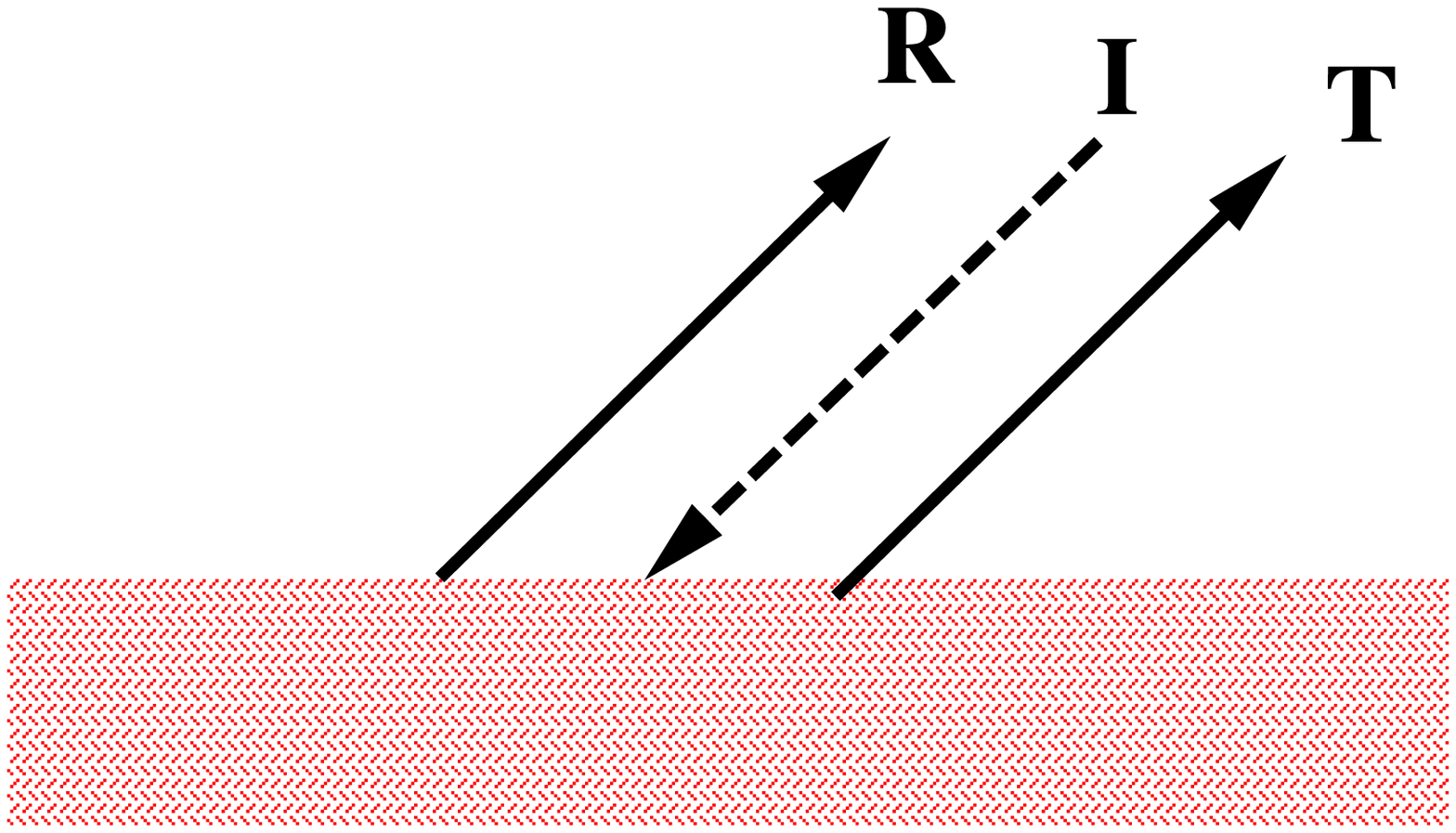}
\label{Figure 3}
\end{center}
\begin{caption}[]

The force due to an incident wave, $I$, is cancelled by the sum 
of the forces due to a reflected wave, $R$, and a transmitted wave, $T$.
\end{caption}
\end{figure}

We next insert the explicit forms for the fields and dipole moment and then 
average the resulting expression over time. The result is
\begin{eqnarray}
F^i = \frac{1}{6} \,A^2\, R\, \Bigl\{ &\alpha_1& \bigl[ 
(k^i - k'^i)({\bf \hat\epsilon}\cdot{\bf \hat\epsilon'}) 
+ 2 \hat\epsilon^i ({\bf k}\cdot{\bf \hat\epsilon'}) 
- 2 \hat\epsilon'^i ({\bf k'}\cdot{\bf \hat\epsilon}) \nonumber \\
&+& 2 \omega\, {\bf \hat\epsilon'}\times({\bf k}\times{\bf \hat\epsilon})
- 2 \omega\, {\bf \hat\epsilon}\times({\bf k'}\times{\bf \hat\epsilon'}) 
  \bigr] \sin\Delta   \nonumber \\
 + &\alpha_2& \bigl[ 
(k^i + k'^i)({\bf \hat\epsilon}\cdot{\bf \hat\epsilon'}) +
2 \hat\epsilon^i ({\bf k}\cdot{\bf \hat\epsilon'}) +
2 \hat\epsilon'^i ({\bf k'}\cdot{\bf \hat\epsilon}) \nonumber \\
&+& 2 \omega\, {\bf \hat\epsilon'}\times({\bf k}\times{\bf \hat\epsilon})
+2 \omega\, {\bf \hat\epsilon}\times({\bf k'}\times{\bf \hat\epsilon'}) 
  \bigr] \cos\Delta \Bigr\} \,. \label{eq:force2}
\end{eqnarray}
Here $\Delta = ({\bf k'} - {\bf k})\cdot {\bf x}_0 + \delta$ is the phase 
difference between the incident and reflected waves at the location of the 
particle.

\begin{figure}
\begin{center}
\leavevmode\epsfysize=5cm\epsffile{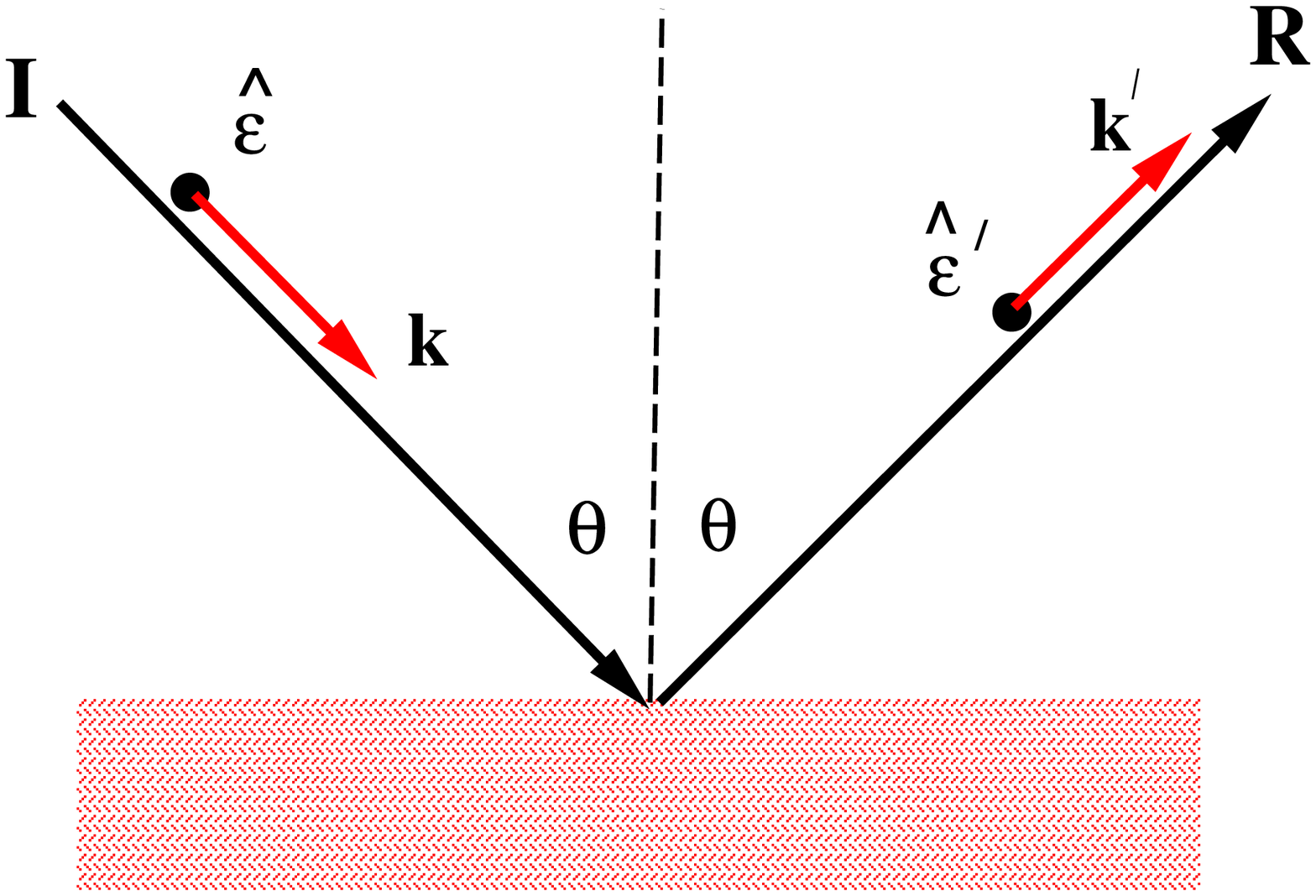}
\label{Figure 4}
\end{center}
\begin{caption}[]

The wavevectors, ${\bf k}$ and ${\bf k'}$, and polarization vectors,
$\hat \epsilon$ and $\hat \epsilon'$, for the incident and reflected 
parts of an S-polarized wave are illustrated.
\end{caption}
\end{figure}

Let us further evaluate this expression. Let the $z$-direction be perpendicular
to the interface and let $\theta$ be the angle of incidence. Then
\begin{equation}
k'_z = - k_z = \omega \, c \,
\end{equation}
where $c = \cos \theta$. Furthermore,
\begin{equation}
\Delta = 2\, k'_z\, z + \delta = 2\, \omega\, z\, c + \delta \,.
\end{equation}
We must now specify the polarization state. We adopt a linear
polarization basis, using the usual S 
(${\bf \hat\epsilon}$ perpendicular to the plane of incidence) and P 
(${\bf \hat\epsilon}$ parallel to the plane of incidence) states.
For S-polarization (Fig. 4), we have
\begin{equation}
{\bf \hat\epsilon'} = {\bf \hat\epsilon}
\end{equation}
and 
\begin{equation}
{\bf \hat\epsilon}\times({\bf k}\times{\bf \hat\epsilon}) = {\bf k}
   \,.
\end{equation}
Only the $z$-component of the force will remain after summation over all
modes; so we need only consider that component. For S-polarization,
we find
\begin{equation}
F_S^z = - A^2\, R_S\, \alpha_1 \, c\, \sin\Delta \,. \label{eq:force_S}
\end{equation}
For P-polarization (Fig. 5), we have that
\begin{equation}
{\bf \hat\epsilon}\cdot{\bf \hat\epsilon'} = \cos 2\theta  \,,
\end{equation}
\begin{equation}
{\bf \hat\epsilon}\cdot{\bf \hat k'}= {\bf \hat\epsilon'}\cdot{\bf \hat k}
= \sin 2\theta  \,,
\end{equation}
\begin{equation}
{\bf \hat\epsilon}\times({\bf k'}\times{\bf \hat\epsilon'}) = -{\bf k} \,,
\end{equation}
\begin{equation}
{\bf \hat\epsilon'}\times({\bf k}\times{\bf \hat\epsilon}) = -{\bf k'} \,,
\end{equation}
and 
\begin{equation}
\epsilon_z = - \epsilon'_z = \sin\theta \,.
\end{equation}
With the aid of these relations, Eq.~(\ref{eq:force2}) can be written for
the case of P-polarization as
\begin{equation}
F_P^z =  A^2\, R_P\, \alpha_1 \, c\,(1 -2c^2)\, \sin\Delta \,.
                                                  \label{eq:force_P}
\end{equation}
Note the force produced by the interference of incident and reflected waves
depends upon $\alpha_1$, the real part of the polarizability, rather than on
the imaginary part as in Eq.~(\ref{eq:single_plane}).

\begin{figure}
\begin{center}
\leavevmode\epsfysize=5cm\epsffile{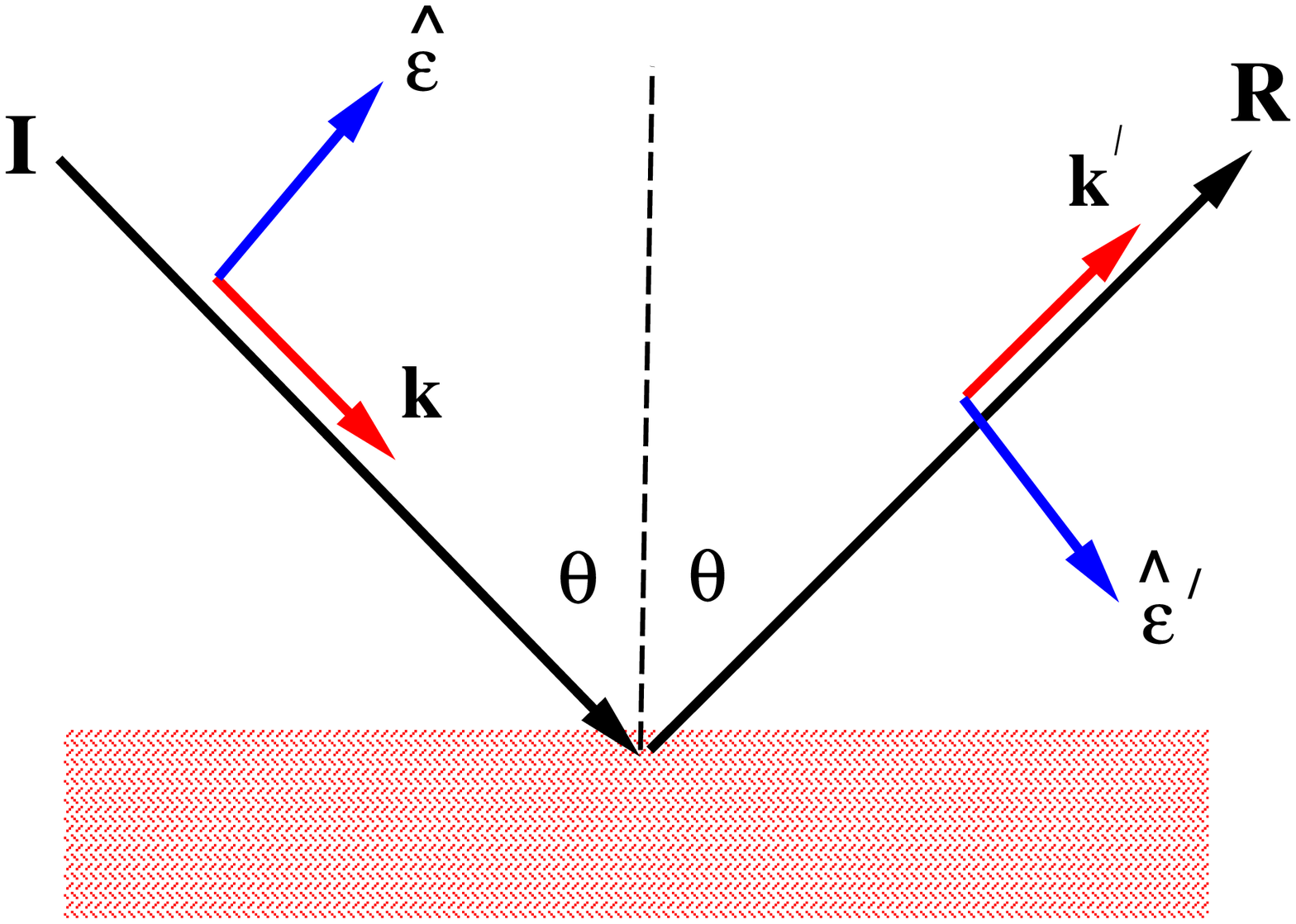}
\label{Figure 5}
\end{center}
\begin{caption}[]

The wavevectors and polarization vectors for a P-polarized 
wave are illustrated. 
\end{caption}
\end{figure}

The net force is obtained by integration of $F_S^z + F_P^z$ over all modes
for which $k_z \leq 0$:
\begin{equation}
F = \int d^3 k \, (F_S^z + F_P^z) =
 2 \pi \int_0^\infty d \omega \, \omega^2 \int_0^1 d c\, (F_S^z + F_P^z)\,.
\end{equation}
The modes are correctly normalized if we set
\begin{equation}
A^2 = \frac{4\pi\, \omega}{ (2\pi)^3}\,.
\end{equation}
This leads to our final result for the force in the direction away from
the interface:
\begin{equation}
F = \frac{1}{\pi} \int_0^\infty d \omega \, \omega^4\, \alpha_1(\omega)\,
\int_0^1 d c\, c\, \Bigl[ -R_S\, \sin(2\, \omega\,z\, c + \delta_S)
+ R_P\, (1-2c^2)\, \sin(2\, \omega\,z\, c + \delta_P) \Bigr] \,. 
                                                     \label{eq:force3}
\end{equation}

It is of interest to note that this result may also be derived from an effective
interaction Hamiltonian of the form of Eq.~(\ref{eq:H_int}), except with
the static polarizability $\alpha_0$ replaced by the real part of the dynamic
polarizability, $\alpha_1(\omega)$.  The interaction potential is given by
first order perturbation theory \cite{TS,milonni} to be
\begin{eqnarray}
V &=& \langle H_{int} \rangle   
= \frac{1}{2 \pi} \int_0^\infty d \omega \, \omega^3\, \alpha_1(\omega)\,
                                                          \nonumber \\
&\times& \int_0^1 d c\, \Bigl[ -R_S\, \cos(2\, \omega\,z\, c + \delta_S)
+ R_P\, (1-2c^2)\, \cos(2\, \omega\,z\, c + \delta_P) \Bigr] \,, \label{eq:pot}
\end{eqnarray}
so that 
\begin{equation}
F = - {\bf \nabla} V \,.
\end{equation}

\section{The Force between a Dielectric Sphere and a Perfectly Conducting Plane}
\label{sec:perfect}

Let us consider the the limit of Eq.~(\ref{eq:force3}) in which the interface
is a perfect conductor. In this limit, we have 
\begin{equation}
R_S = R_P = 1\, , \label{eq:Rperfect}
\end{equation}
and 
\begin{equation}
\delta_S = \delta_P = \pi \,. \label{eq:deltaperfect} 
\end{equation}
This leads to 
\begin{eqnarray}
F &=& \frac{2}{\pi} \int_0^\infty d \omega \, \omega^4\, \alpha_1(\omega)\,
\int_0^1 d c\, c^3\, \sin(2\, \omega\,z\, c)   \nonumber \\
&=&  - \frac{1}{4\pi z^4} \int_0^\infty d \omega \, \alpha_1(\omega)\,
                                                 \nonumber \\  
&\times&     \Bigl[ 3\sin 2\omega z -6 z\,\omega\, \cos 2\omega z
 -6 z^2\,\omega^2\,\sin 2\omega z
  + 4z^3\,\omega^3\,\cos 2\omega z \Bigr] \,.  \label{eq:force4}
\end{eqnarray}    
Note that there are no evanescent modes in this case; so the previous 
approximation of ignoring such modes is not needed here.

Now consider a sphere of radius $a$ composed of a uniform material with
dielectric function $\varepsilon(\omega)$. The complex polarizability is given
by
\begin{equation}
\alpha(\omega) = a^3\, \frac{\varepsilon(\omega) - 1}{\varepsilon(\omega) +2}
                              \,.  \label{eq:alpha}
\end{equation}
We will take the dielectric function to be that of the Drude model,
\begin{equation}
\varepsilon(\omega) = 1 - \frac{\omega_p^2}{\omega(\omega + i\gamma)} \,,
                                                \label{eq:epsilon}
\end{equation}
where $\omega_p$ is the plasma frequency and $\gamma$ is the damping parameter.
From Eqs.~(\ref{eq:alpha}) and (\ref{eq:epsilon}), we find that the real part of 
the polarizability is given by
\begin{equation}
\alpha_1 =  a^3\, \omega_p^2\, 
\frac{\omega_p^2 -3 \omega^2}{(3 \omega^2 -\omega_p^2)^2 + 9 \omega^2 \gamma^2}
                               \,.  \label{eq:alpha_1}
\end{equation}
Note that although $\alpha(\omega)$ has poles only in the lower half-$\omega$
plane, its real part, $\alpha_1(\omega)$, has poles in both the upper and lower
half planes. 

If we insert Eq.~(\ref{eq:alpha_1}) into  Eq.~(\ref{eq:force4}), we must 
evaluate the following set of integrals:
\begin{equation}
I_1 = \int_0^\infty d \omega \, \alpha_1(\omega)\, \sin(2\, \omega\, z)
= Im \int_0^\infty d \omega \, \alpha_1(\omega)\, e^{2\,i\, \omega\,z} \,,
                                                          \label{eq:I1}
\end{equation}
\begin{equation}
I_2 = \frac{1}{2} \frac{d I_1}{d z} =
\int_0^\infty d \omega \, \alpha_1(\omega)\, \omega\, \cos 2\omega z
                                                \,, \label{eq:I2}
\end{equation}
\begin{equation}
I_3 = -\frac{1}{2} \frac{d I_2}{d z} =
\int_0^\infty d \omega \, \alpha_1(\omega)\, \omega^2\, \sin 2\omega z
                                                \,, \label{eq:I3}
\end{equation}
and
\begin{equation}
I_4 = \frac{1}{2} \frac{d I_3}{d z} =
\int_0^\infty d \omega \, \alpha_1(\omega)\, \omega^3\, \cos 2\omega z
                                                \,, \label{eq:I4}
\end{equation}
In terms of these integrals, the force between the sphere and the plate
is
\begin{equation}
F = - \frac{1}{4\pi z^4}\, \Bigl( 3\, I_1 -6\,z\,I_2 -6\,z^2\,I_3
                                    + 4 \,z^3\,I_4 \Bigl) \,.
\end{equation}
The second integral in Eq.~(\ref{eq:I1}) may be evaluated by rotating the 
contour of integration to the positive imaginary axis (Fig. 6). 
However, in this process we will also acquire
a contribution from the residue of the pole of $\alpha_1(\omega)$ at
$\omega = \Omega + \frac{1}{2}\, i\, \gamma$, where 
 \begin{equation}
\Omega = \frac{1}{6} \sqrt{12 \omega_p^2 - 9 \gamma^2} \,.
\end{equation}
The result may be written as
 \begin{equation}
I_1 = J_1 + P_1\,.
\end{equation}
Here integrating over imaginary frequency yields
\begin{equation}
J_1 = \int_0^\infty d \xi \, \alpha_1(i\xi)\, e^{-2\, \xi\, z}
= a^3\, \omega_p^2\, \int_0^\infty d \xi \,
\frac{3 \xi^2 +\omega_p^2}{(3 \xi^2 +\omega_p^2)^2 - 9 \xi^2 \gamma^2} \,
e^{-2 z \xi} \,,
\end{equation}
and the residue of the pole is
\begin{equation}
 P_1 = - \frac{\pi\,a^3\, \omega_p^2}{6\, \Omega}\, e^{-\gamma z} \,
         \cos 2\Omega z \,.
\end{equation}

\begin{figure}
\begin{center}
\leavevmode\epsfysize=5cm\epsffile{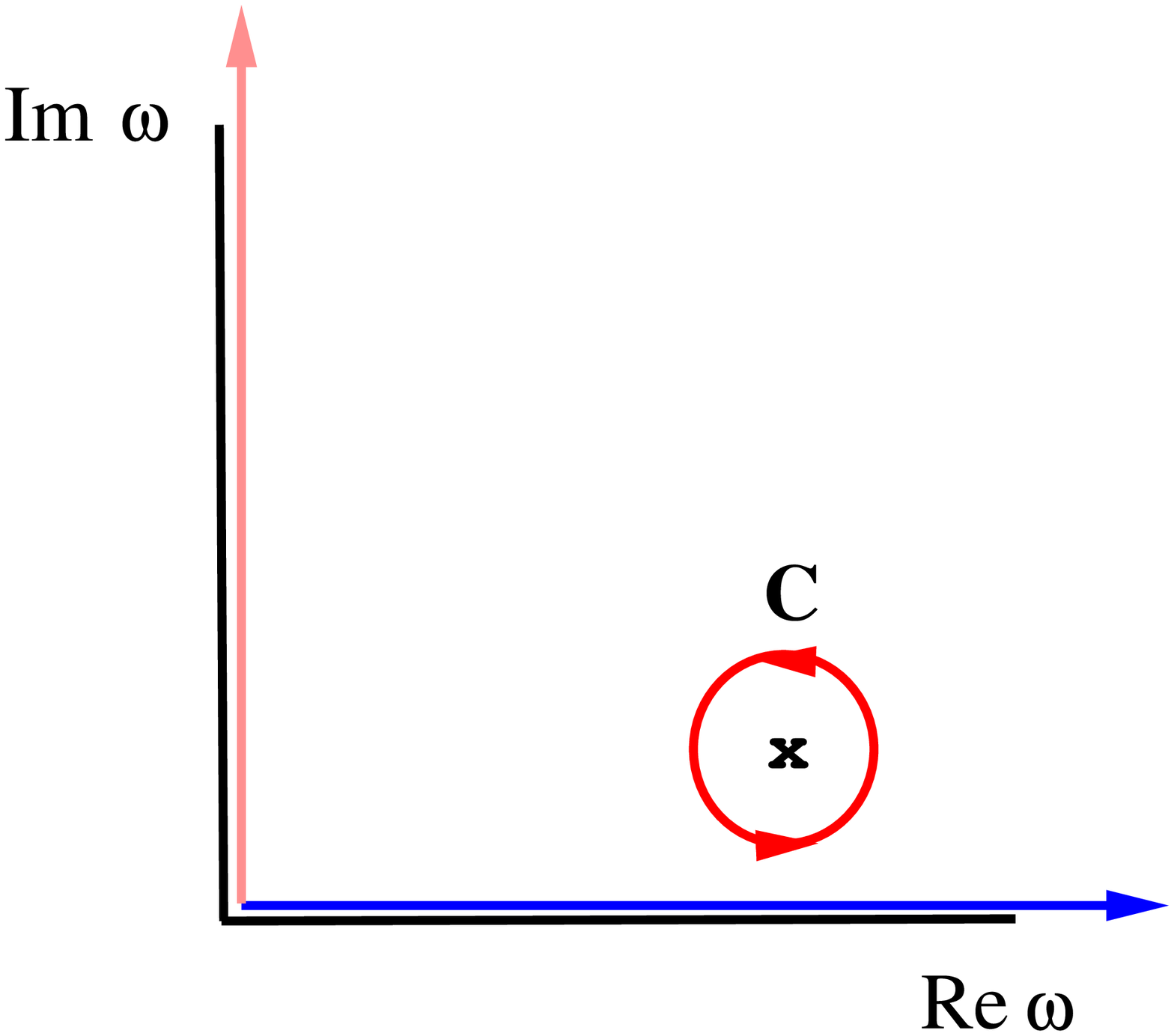}
\label{Figure 6}
\end{center}
\begin{caption}[]

The contours of integration for integrals of the form of
Eq.~(\ref{eq:I1}) are illustrated. The integral on real $\omega$ can
be expressed as a sum of an integral on imaginary $\omega$, plus a 
contribution, $C$, coming from the pole at 
$\omega=\Omega + \frac{1}{2} i \gamma$. 
\end{caption}
\end{figure}

These results may be combined to obtain our final expression for the force
between the sphere and the plate, which may be written as
\begin{equation}
F = J + P \,,
\end{equation}
where $J$ is the net contribution from integrals along the imaginary axis, and
$P$ is that from the pole at $\omega = \Omega + \frac{1}{2}\, i\, \gamma$.
The explicit forms of these two contributions are
\begin{equation}
J = - \frac{a^3\, \omega_p^2}{4 \pi \, z^4}\, \int_0^\infty d \xi \,
\frac{(3 \xi^2 +\omega_p^2)(4 z^3 \xi^3 +6z^2 \xi^2 +6z\xi +3)}
     {(3 \xi^2 +\omega_p^2)^2 - 9 \xi^2 \gamma^2}\;  e^{-2 z \xi} \,,
                                                        \label{eq:J}
\end{equation}
and
\begin{eqnarray}
P &=& - \frac{a^3\, \omega_p^2}{48 \,\Omega\, z^4}\, e^{-\gamma z} \,
\Bigl[ 2\Omega \,z\,(4\,\Omega^2\, z^2 -
   3\gamma^2\, z^2 -6\gamma z -6) \sin 2\Omega z              \nonumber \\
 &+& (12\gamma \,\Omega^2\, z^3 -\gamma^3 \,z^3 +12\Omega^2\, z^2 
-3\gamma^2\, z^2 -6\gamma \,z -6) \cos 2\Omega z \Bigr] \,.  \label{eq:P}
\end{eqnarray}
(Here and at other points in this paper, the calculations were performed
with the aid of the symbolic algebra program MACSYMA.)

In the case that $\gamma =0$, the integral for $J$ may be evaluated in terms
of sine and cosine integral functions. In the limit of small separations,
one finds for this case that
\begin{equation}
J \sim a^3\, \omega_p\, \left( -\frac{\sqrt{3}}{8 z^4} + 
\frac{\omega_p}{6\, \pi\, z^3} +O(z^{-1}) \right)
\end{equation}
and that
\begin{equation}
P \sim a^3\, \omega_p\, \left( \frac{\sqrt{3}}{8 z^4} + O(z^{0}) \right)\,.
\end{equation}
Thus the leading terms cancel, and we find a repulsive force in this limit:
\begin{equation}
F \sim \frac{a^3\, \omega_p^2}{6\, \pi\, z^3} +O(z^{-1})\,, \qquad
                             a \ll z \ll \omega_p^{-1} \,.
\end{equation}

It is of particular interest that $P$ contributes an oscillatory term to the 
force. In the large separation limit, $z \gg 1/\omega_p$, we have that
\begin{equation}
J \sim - \frac{3\, a^3}{2 \pi\, z^5} \,.  \label{eq:CPforce}
\end{equation}
This is just the attractive force due to the asymptotic Casimir-Polder 
potential, Eq. ~(\ref{eq:CP}), where $\alpha_0 = a^3$ is the static
polarizability of the sphere. The oscillatory term becomes, in the large 
distance limit,
\begin{equation}
P \sim - \frac{\Omega\, \omega_p^2\, a^3}{12\, z} \; e^{-\gamma\, z} \; 
\Bigl( 2\Omega\, \sin 2\Omega z + 3\gamma\, \cos 2\Omega z \Bigr)\,.
                                                \label{eq:large_z}
\end{equation}
Although this term is exponentially decaying, it is possible for it
still to be significant in the asymptotic region if, as is typically the case,
$\gamma \ll \omega_p$. In this case, the oscillatory term $P$ will dominate
the Casimir-Polder term, $J$, and lead to a series of stable equilibrium
points at finite distance from the boundary, separated from one another by a
distance of approximately $\ell = \pi/\Omega$. A plot of the force at various
separations is given in Fig. 7. 

\begin{figure}
\begin{center}
\leavevmode\epsfysize=8cm\epsffile{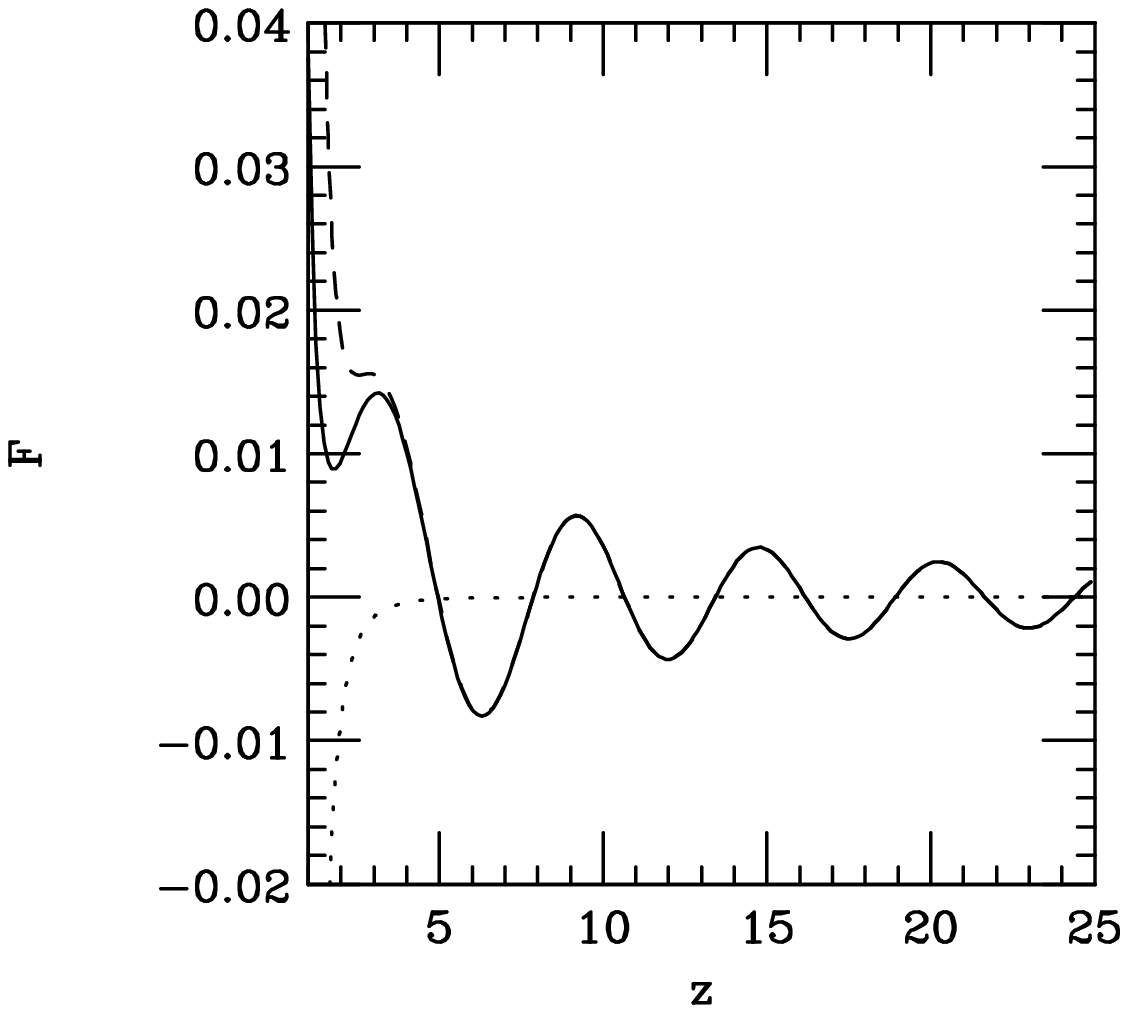}
\label{Figure 7 }
\end{center}
\begin{caption}[]

The force $F$ between a sphere and a perfectly reflecting wall
is illustrated in the case where $\gamma = 0.005 \omega_p$, with  $F$ 
in units of $\omega_p^5 a^3$ and $z$ in units of $\omega_p^{-1}$. The stable
equilibrium points are the zeros of $F$ where the slope is negative.
Here $F >0$ corresponds to repulsion. The dotted line is the contribution
of $J$, the imaginary frequency integral Eq.~(\ref{eq:J}), and the dashed
line is that of $P$, the pole contribution Eq.~(\ref{eq:P}).
\end{caption}
\end{figure}

One might imagine trying to levitate the spheres in the Earth's gravitational
field by this means. This will occur if $F_{max} \geq F_g$, where $F_{max}$
is  $F$ evaluated at a peak value, and $F_g$ is the force of gravity.
The ratio of these two forces may be expressed as
 \begin{equation}
\frac{F_{max}}{F_g} \approx 27\, \left(\frac{\omega_p}{1 eV}\right)^4\,
\left(\frac{1 \mu m}{z}\right)\, \left(\frac{1 g/cm^3}{\rho}\right) \,
e^{-5\,(\gamma/ 1eV)\,(z/ 1 \mu m)} \,,
\end{equation}
where $\rho$ is the mass density of the sphere. We have assumed that
$\gamma \ll \omega_p$, so   $\Omega \approx \sqrt{3}\,\omega_p/3$.
Let $z = z_c$ be the distance at which this ratio of forces is unity, and hence
the maximum distance above the interface at which levitation can occur.
In Table 1,  values of $z_c$ for various alkali metals are given, along with
appropriate input parameters.

\begin{table}
\begin{tabular}{|l|c|c|c|c|c|}  \hline
 & $\rho$ & $\omega_p$ & $\gamma$ & $\ell$ &  $z_c$ \\   \cline{2-6}
Li & 0.53 & 6.6 & 0.031 & 0.16 & 49 \\
Na & 0.97 & 5.6 & 0.028 & 0.19 & 46 \\
K &  0.86 & 3.8 & 0.021 & 0.28 & 47 \\  \hline
\end{tabular}
  \caption{Parameters for some alkali metals. The Drude model parameters,
$\omega_p$ and $\gamma$, taken from Ref. \cite{Palik}  , are in eV. The
maximum levitation height, $z_c$, and the separation between equilibrium
points, $\ell$, are in $\mu m$.}
\end{table}

The maximum elevation $z_c$ at which a sphere could levitate is in the range
of $46\, -\, 49 \mu m$. This is rather larger than the distance
at which Casimir forces are usually expected to have a noticeable effect.
Recall that all of the discussion in this paper is at zero
temperature. Thermal effects at finite temperature can mask this
vacuum energy effect. For example, for a sodium sphere of radius $a = 50 nm$
near the maximum levitation height, the difference in potential energy between
successive equilibrium points corresponds to a temperature of approximately
$0.1 K$, and would be observable only at low temperatures. On the other hand,
the corresponding energy difference near the minimum levitation height is
about $2000 K$. Thus the first several equilibrium points should be observable 
at room temperature. The use of a perfectly reflecting wall should be a
reasonable approximation so long as the plasma frequency of the material
in the wall is large compared to that in the sphere. Thus, a wall composed
of aluminum ($\omega_p = 14.8 eV$)  \cite{Palik2}
 is a good reflector 
at frequencies of the order of the plasma frequencies of the alkali metals.

\section{Discussion}
\label{sec:final}

In the previous sections, we have seen that a polarizable sphere with a
dispersive polarizability in the vicinity of a perfectly reflecting boundary can
experience a Casimir force which is much larger than would be experienced by a 
perfectly conducting sphere at the same separation. This can be understood
in terms of the oscillatory frequency spectrum of vacuum energy effects.
Cancellations between different parts of the spectrum which occur in the
perfectly conducting limit seem to be upset by the dispersive properties of
the sphere's material. A perfectly reflecting sphere would have
a frequency independent polarizability of $\alpha =\alpha_0 = a^3$, and
the force exerted by the wall would be given by Eq.~(\ref{eq:CPforce})
at all separations. In addition to its amplification, the force now becomes 
an approximately
oscillatory function of position, leading to the possibility of trapping
the sphere in stable equilibrium. 

Note that this type of oscillatory force does not arise in the case of
a pair of half-spaces of dielectric material separated by a gap. If the
material in the half-spaces is a homogeneous dielectric, whose dielectric
function satisfies the Kramers-Kronig relations, then 
the Lifshitz theory \cite{Lif}
predicts a force of attraction which is always less than that in the case
of two perfectly conducting planes. Apparently, the effect of the infinite
spatial volume of the  half-spaces is to average over the spatial
oscillations. A similar result was found recently by  Lambrecht {\it et al.}
\cite{LJR} for the case of mirrors for a scalar field in one spatial
dimension.

It is of interest to compare the macroscopic sphere problem discussed in
this paper with the problem of an atom near a perfect mirror. The case
where the atom is in the ground state was discussed in the original
Casimir-Polder paper \cite{CP}, where a monotonically decreasing expression was
obtained which reduces to Eq.~(\ref{eq:CP}) in the large $z$ limit.
This result is of the same form as the contribution $J$ to the net force
found in Section~\ref{sec:interface} coming from the integration over 
imaginary frequencies. Various authors \cite{TS,milonni,SDM} have treated
the problem of a polarizable particle near an interface. However, these authors 
were primarily interested in the case where the polarizable particle is an
atom in its ground state, and hence included only the imaginary frequency
contribution.
 The case of an atom in an excited state was treated by Barton 
\cite{Barton} and other 
authors \cite{CPS,WS,HS}, who found that the potential now
has an oscillatory component. Furthermore, this oscillatory term at large
distances has a form similar to Eq.~(\ref{eq:large_z}), with the magnitude of
the oscillatory part decreasing as $1/z$. Thus at large separations, the net
potential is dominated by this oscillatory term. In the case of the atom
in an excited state, the oscillating potential can be given a classical
interpretation: The atom behaves like a radiating antenna in the presence
of a mirror. Such an antenna will experience an oscillatory backreaction
force whose sign depends upon whether the reflected wave interferes constructively
or destructively with the original radiated wave. The oscillatory force
found in the present problem does not seem to have such an interpretation,
because the dielectric sphere is not radiating. Nonetheless, it is plausible 
that there should be some parallels between an atom in an excited state
and a macroscopic system such as the sphere with a continuum of quantum
states just above the ground state.

The oscillatory force can be understood in this case as arising from 
a position dependence of the cancellation of the different parts of the
frequency spectrum. One can see from Fig. 1 that a particle whose 
polarizability is nonzero only in a narrow band of frequency will experience
an oscillatory force. (See Ref. \cite{Ford93} for further discussion of this
point.) The delicate cancellation is perhaps one reason that it is difficult
to predict the sign of a Casimir force in advance of an explicit calculation.

Finally, let us recall the assumptions which were employed in the analysis of 
this paper. The general formula for the force, Eq.~(\ref{eq:force3}), was 
derived in Sect.~\ref{sec:interface} assuming the scattered wave is dipole
and that there are no evanescent modes. The dipole approximation should be
valid so long as the size of the particle is small compared to the wavelength
of any modes which contribute significantly to Eq.~(\ref{eq:force3}). The
assumption of no evanescent modes places some restrictions on the material 
of the interface. In particular, a perfectly conducting interface will have
no evanescent modes. More generally, in frequency ranges in which the real part 
of the index of refracion is less than unity, there will be no such modes.
This will be the case for all frequencies if the interface is composed of
a metal for which the collisionless Drude model (Eq.~(\ref{eq:epsilon})
with $\gamma =0$) is a good approximation. In Sect.~\ref{sec:perfect}, we
made some further approximations. These included the assumption that the
particle is a small sphere whose dielectric function has the form given
by the Drude model, Eq.~(\ref{eq:epsilon}). Here the dipole approximation
is expected to be valid when $a \ll \omega_p^{-1}$. A final approximation 
was made in assuming that the interface is perfectly conducting. This is
expected to be valid when the interface is composed of a metal
whose plasma frequency is large compared to that of the sphere. Then the
dominant contributions to Eq.~(\ref{eq:force3}), those for which $\alpha_1
\not= 0$, come from modes for which
Eqs.~(\ref{eq:Rperfect}) and (\ref{eq:deltaperfect}) are approximately valid.
The extension of the results of this paper to the case where the interface
is an imperfect reflector is currently under investigation.

\vskip .8cm
{\bf Acknowledgments:}  I would like to thank G. Barton, T. Jacobson, 
P.W. Milonni,
V. Sopova, and L. Spruch for useful conversations. This work was
supported in part by the National Science Foundation (Grant No. PHY-9507351).

\end{document}